\documentclass[12pt,onecolumn,draftcls]{IEEEtran} 

\usepackage{verbatim}
\usepackage{amssymb}
\usepackage{amscd}
\usepackage{amsmath}
\usepackage{graphicx}
\usepackage{epstopdf}
\usepackage{epsfig}
\usepackage{amsthm}

\newtheorem{theorem}{Theorem}[section]

\newtheorem{proposition}{Proposition}[section]
\newtheorem{lemma}{Lemma}[section]
\newtheorem{definition}{Definition}[section]

\renewcommand{\thesection}{\arabic{section}}
\renewcommand{\theequation}{\arabic{section}.\arabic{equation}}
\renewcommand{\thesubsection}{\thesection.\arabic{subsection}}

\begin{document}
\title{Transmission Sequence Design and Allocation for Wide Area Ad Hoc Networks
\author{Wing Shing Wong, {\em Fellow, IEEE}\\
Department of Information Engineering\\
The Chinese University of Hong Kong}
\thanks{Email: wswong@ie.cuhk.edu.hk, personal website: http://www.ie.cuhk.edu.hk/people/wing2.shtml.
The author would like to acknowledge support from the National Natural Science Foundation of China under grant number 61174060 and Shenzhen matching grant GJHS20120702105523301.}}
\IEEEaftertitletext{\vspace{-2\baselineskip}}
\maketitle

\begin{abstract}
In this paper we examine the problem of designing and allocating transmission sequences to users in a mobile ad hoc network that has
no spatial boundary.   A basic tenet of the transmission sequence approach for addressing media access control is that under normal operating conditions, there is no feedback triggered re-transmission.   This obviously is a major departure from the Slotted-ALOHA or CSMA type approaches.  While these solutions enjoy excellent throughput performance, a fundamental drawback is that they are based on feedback
information.   For systems without naturally defined central controller that can play the role of a base station, the task of providing feedback information could easily become unmanageable.  
This highlights the advantage of the feedback-free  approach.  A second advantage is the ability to handle unlimited spatial coverage.  We propose in this paper a concept for media access control that is akin to frequency reuse.   However, instead of reusing frequency, the new approach allows transmission sequences be reused.   A study of the transmission sequence approach against other approaches is conducted by comparing the minimal
frame lengths that can guarantee the existence of conflict-free transmissions.
\end{abstract}


\section{INTRODUCTION}\setcounter{equation}{0}
One of the basic challenges of wireless ad hoc network design is to ensure efficient simultaneous access to the shared communication media for massively many, geographically distributed users.   For wide area multi-hop ad hoc networks the underlying issues are even more demanding.   

Traditional multiple access control mechanisms are either frequency-division based (FDMA), code-division based (CDMA),
or time-division based (TDMA).  Typical FDMA or CDMA schemes contain elaborate steps
to ensure a transmitter and its intended receiver are synchronized to the same frequency band or same spreading code.
These steps are necessary because it is physically impossible to ask the receiver to monitor all transmission channels or the whole code-space
at the same time.  Instead, the receiver and transmitter must find a way to match each other.
The matching algorithm usually requires some form of hand-shaking or acknowledgment message exchanges and is costly. 
For ad hoc networks supporting datagram services, these connection or synchronization setup procedures may prove to be
complicated and inefficient.  This motivates the search for protocols with simple set-up requirements.

For the TDMA approach, a frequency channel is divided into contagious time slots of fixed duration.  These slots are usually grouped into periodic frames.
While it is relatively simple for transceivers to monitor all the time slots, contention will occur if multiple users
decide to broadcast at the same time.  Hence, some TDMA schemes also require slot allocation by a centralized controller
which will incur connection setup costs.   Nevertheless, setup-free time-division based schemes are also available,
such as Slotted-ALOHA or  Carrier Sense Multiple Access (CSMA).
These schemes are efficient but require channel information feedback for contention resolution.
While feedback can be readily provided on systems with base-stations, it poses cumbersome burdens for ad hoc networks without a differentiated control management hierarchy.

In this paper, we develop a setup-free approach based on two closely-related concepts, namely
{\em code-based scheduling} and {\em protocol sequence}.   Both of these concepts are feedback-free.
That is, under normal operating conditions, there is no need to feedback channel status information to the users and therefore
there is no need to identify a centralized controller to broadcast the channel state.  Both of these approaches define the
access protocol by using binary {\em transmission sequences} to indicate permission to transmit on a particular time slot or
to remain silent.  We follow here the convention that a ``one" represents permission to transmit and a ``zero" to remain silent.
The issue of defining an orderly and efficient access to the channel is then reduced to a question of designing transmission sequences
that satisfy the following conditions:
\begin{enumerate}
\item System performance should achieve targeted objectives.
\item There should be a mechanism for users to acquire their transmission sequences automatically.
\item The sequence parameters should allow various traffic scenarios be readily accommodated.
\end{enumerate}

The statement of the first task is deliberately vague since there can be a variety of system goals to be achieved, such as maximization of
system throughput, minimization of average delay, minimization of maximum delay, and so on.  In this study, we focus
on the objective of providing service guarantee that each user can broadcast at least one  contention-free packet to its
neighbors in each frame and that the frame length is minimized.
However, our investigation can be extended to throughput maximization, for example.

No matter what the system objective is, the optimal solution would depend on the degree of time synchronization among users.
Since Global Navigation Satellite Systems (GNSS) are ubiquitous these days and they can provide time synchronization,
it is natural to assume that users have access to one and their clocks are synchronized up to some intrinsic inaccuracies. 
However, due to propagation delay, it is not possible to perfectly synchronize the clocks among a transmitter
and all of its receivers in an ad hoc network. 
Therefore we need to discuss Media Access Control (MAC) solutions for wide area ad hoc networks under various synchronization conditions.  Details will be described in subsequent sections.

For code-based scheduling, a simple approach for sequence distribution is to pre-assign them to all
users.   This assumes the maximum number of users in the network is unchanging and {\em a priori} known.
While there are other alternatives for automatic transmission sequence allocation, (see for example \cite{WuYi1},)
in this paper we propose a scheme that is based on the geographic location of the distributed users.
A key premise is that all users have access to their location information at marked time instances through a Global Navigation Satellite System (GNSS) such as the Global Positioning System (GPS).   In order to define the allocation scheme,
we tessellate the geographic region to be covered into regular hexagonal quantization cells.
The cell boundaries are assumed to be fixed and known to all users.  Therefore, all users can identify themselves
with a location cell at any time instance marked by a GNSS signal broadcast.  The transmission sequences are pre-assigned to the
location cells and then dynamically assigned to the users depending on their location.

A distinct feature of our allocation scheme is that it takes advantage of the
idea that users separated by distance larger than the hearing range can reuse the same time slots and hence can be assigned
the same transmission sequence.   This form of {\em sequence reuse} reminds us of,
and yet is distinct from, frequency-reuse in cellular networks.

Since the methodology of our solution approach comes from topology transparent code-based scheduling
and protocol sequence, it would be appropriate to recall basic ideas from these topics.

Starting from \cite{Solomon}, various researchers have proposed using code-based scheduling 
to address the media access problem.   The sequences defined in \cite{CF94}, \cite{RK2}, and \cite{RK3}, for example,
can be applied to the multi-hop ad hoc networks under our consideration.
By means of various coding schemes, such as Reed-Solomon (RS) codes, one can construct large sequence spaces to support the stated applications.   However, a major issue is that these sequences require perfect synchronization.
In an ad hoc network, even if transmitters are perfectly synchronized
to a global clock, the times when packet are received are typically not, due to propagation delays.

Protocol sequences are more suitable for ad hoc network applications.
In \cite{Massey82} and \cite{Massey85},
Massey and his co-authors proposed using binary sequences to address the MAC problem for asynchronous users in a cell
and coined the nomenclature.
Simply put, protocol sequences are binary sequences with a common period and
designed with special Hamming cross-correlation properties.
These sequences are expected to be used repeatedly in real applications.
For asynchronous systems, the starting points of users are indeterminate,
hence cyclic shifts of a particular sequence should not be treated as distinct codes.
Thus, protocol sequences are closely related to the {\em cyclically permutable codes} defined by Gilbert \cite{Gilbert63}.
These are binary block codes with each codeword having a cyclic order that  is
equal to the block length.
(A cyclic order is the number of shifts required to return the resulting sequence to its original form.)
In \cite{Nguyen92}, various methods for constructing cyclically permutable codes were proposed.
These constructions were based on a classical application of the Chinese Remainder Theorem (CRT) which defines a one-to-one, onto
correspondence between vectors and matrices.  Subsequently, the author and his coworkers also used the CRT correspondence
to construct new classes of protocol sequences that are known as CRT sequences (\cite{SW-ISIT}, \cite{SETA10}, and \cite{CRT}). These
CRT sequences are closely related to the {\em Generalized Prime Sequences} defined in \cite{WuYi1} which in turn are generalizations
of prime sequences \cite{Shaar84} and extended prime sequences \cite{Yang02}.
We will show that cyclically permutable codes are excellent candidates for MAC application in ad hoc networks.

In the following sections, we describe how to design transmission sequences for wide area ad hoc networks under various synchronization conditions.  The GNSS-based solution is presented. 
This approach is based on transmission sequence reuse,
a new concept modeled after frequency reuse.  We also discuss the benefits of
GNSS-based allocation.  Among them is the ability to allow the maximum number of local users to vary according to location.
We propose new sequence constructions to achieve this feature.  The construction is based on the concept
of {\em product sequences} which also relies on the CRT correspondence.

The organization of the rest of the paper is as follows.  The basic model and key assumptions are presented in Section 2.
In Section 3, we recall relevant results and concepts from protocol sequences.
Section 4 describes a new sequence allocation algorithm, the GNSS-based allocation,
that relies on the geographical location of the users.
In Section 5 we present different transmission sequence sets that can be used
in a GNSS-based allocation, including TDMA, code-based scheduling, and protocol
sequences.
In Section 6 we discuss the difference between the
GNSS-based code allocation with traditional code-based scheduling. 
We also describe a new code allocation scheme that can adapt to geographically changing user density.
Some concluding remarks are offered in Section 7.

\section{MULTI-HOP AD HOC NETWORK}\setcounter{equation}{0}
A mobile multi-hop ad hoc network is a network without hierarchical infrastructure in which mobile users can
broadcast data bursts to all their neighbors who are within hearing range \cite{Broch}.
Such type of models is appropriate for describing networks for vehicular applications,
(see for example \cite{Menouar, WuYi1}.)
It is difficult to achieve efficient MAC design using feedback-type protocols
for these networks
since they contain massively many access points which render standard contention resolution algorithms cumbersome.  Thus, feedback-free approaches are investigated
in this paper.

Consider a multi-hop ad hoc network covering a geographic area without bounds,
and hence it can support in theory an unbounded number of users.
All users are mobile, with maximum mobility speed given by $v$ m/s.
All users are continually active in the sense they are always ready to transmit and to receive messages.  Such type of traffic data model is suitable for applications such as neighboring vehicle detection or collision warning in a
{\em Vehicular Ad Hoc Network} (VANET).

All users share a single, slotted frequency channel for duplex communication.
That is, the channel is divided into time slots and at each time slot a mobile user can either transmit or receive signal but cannot do both. 
The time slots are all $\tau$ second long.
A user coordinates transmission and reception in each frame by means of an assigned transmission sequence which has a period
equal to $L$.
This number also defines the number of slots in a {\em frame}.
Unlike code-based scheduling approaches which allocate sequences to users directly, we consider an alternative approach in
which sequences are first allocated to geographic cells, and then to the users through their cell location as explained in
Section 4.

We assume all users have access to a GNSS which broadcasts signals
at marked time instances, $(S_0, S_1, S_2, \ldots )$, where $S_i=iT$ for some $T$. 
The time between any two consecutive marked instances is organized into
a superframe consisting
of $F$ frames, each of which contains $L$ time slots of length $\tau$.   So that
\begin{equation}
T = FL\tau + T_G,
\end{equation}
where $T_G$ stands for the superframe guard time which is added to ensure all packets transmitted in a superframe
will be received during the same superframe.  An upper bound for $T_G$ will be shown later.
We refer to all frames other than the first and last one as {\em normal frames}.  Our analysis will focus on
normal frames to avoid complexities due to boundary efforts.  For large values of $F$,
the boundary effects are negligible.

As common in the analysis of ad hoc networks, we adopt the following assumption.

\noindent
{\bf The Interference Assumption:}
Any receiver can only listen to
those transmitters within a distance $R$ from it, transmissions from location
at a distance equal to or greater than $R$ are simply ignored.
Since we are mainly interested in scenarios where the value of $T$ is in the order of seconds,
we assume the interference relations among users
are unchanged during a superframe, $[S_i, S_{i+1})$, to simplify the analysis.

For proper protocol operation, users are required to synchronize to the global GNSS clock by aligning to
the marked instances -- the exact start time of the $k$ superframe is announced by the GNSS at $S_k$.   However, due to differences
in propagation delay and processing time, actual start times for different users may differ. 
If $t_i$ represents the start time of a superframe for user $i$,
we assume that there is a system-wide integer bound, $\Delta_C$, such that
\begin{equation}
0 \leq t_i - S_i \leq \tau \Delta_C.
\end{equation}
Let
\begin{equation}
\Delta_P = \left \lceil \frac{R}{c\tau}  \right \rceil,
\end{equation}
where $c$ is the speed of light.  Define
\begin{equation}
\Delta = \Delta_C + \Delta_P.
\end{equation}
We assume that
\begin{equation}
\Delta \leq L.
\end{equation}
It follows that if user $a$ transmits a packet at a slot in a normal frame $i$ and the packet is received by user $b$ during frame
$j$ then the difference between $i$ and $j$ is bounded by 1.  That is,
\begin{equation}
|i-j| \leq 1.
\end{equation}
See Figure 1 for illustration.

\begin{figure}
\begin{center}
\includegraphics[width=4.5in]{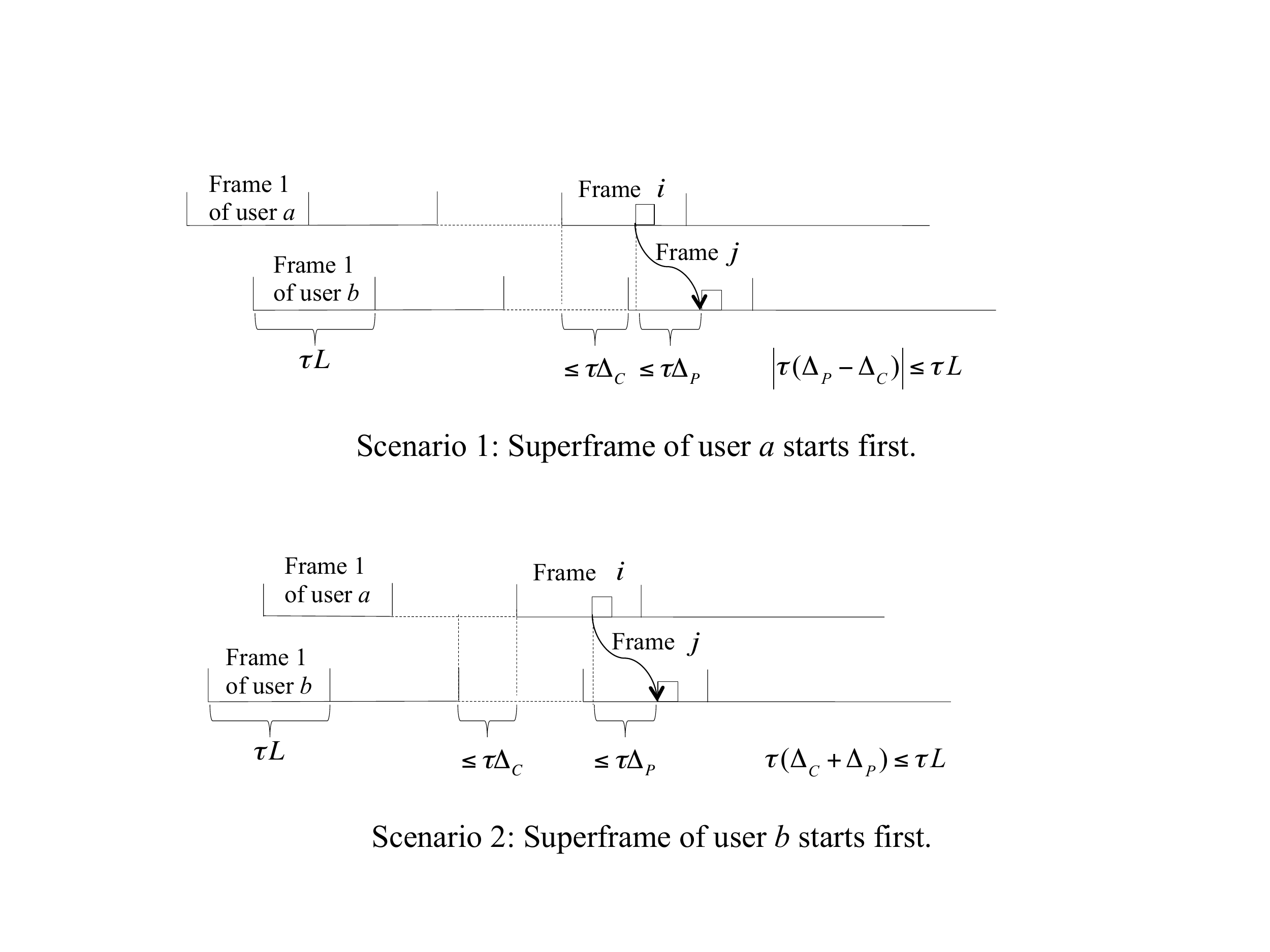}

Figure 1:  Two scenarios for calculating time differences between users
\label{fig1}
\end{center}
\end{figure}
It follows that that the superframe guard time can be set to be $\tau \Delta$.

In our analysis, we assume different types of synchronization, which are defined in terms of the
value of $\Delta$ and $\Delta_C$.
\begin{enumerate}
\item The system is {\em perfectly clock synchronized} if $\Delta_C=0$.
\item The system is {\em perfectly synchronized} if $\Delta=0$.
\item The system is {\em partially synchronized} to $\Delta_0$ slots if $\Delta = \Delta_0 < L$; the system is {\em asynchronous} if $\Delta = L$. 
\end{enumerate}
It is worthwhile to point out that even though these definitions are expressed in terms of a multiple number of slots, the
time differences can be any positive real value bounded by $\tau \Delta$.
In subsequent discussion, we also need to refer to the concept of {\em slot-synchronized}; this refers to
cases where the time difference for any arbitrary transmitter-receiver pair is idealized to an integral multiple of $\tau$
for simplicity in discussion.

Inaccuracies in information are not restricted to time and may also occur in the location coordinates of the users.
These inaccuracies are taken care of by means of quantization of location coordinates into regular hexagonal tiling cells with cell radius $h$ m.    To be more precise, consider a planar coordinate system with the
axes intersecting at a sixty degree angle instead of orthogonally.
We can define a regular tiling of the plane by hexagonal cells with centers
$(md,nd)$ for all integers $m$ and $n$, where $d=\sqrt{3}h$
as shown in Figure 2.  We call these cells {\em quantization cells}.

\begin{figure}
\begin{center}
\includegraphics[width=4.5in]{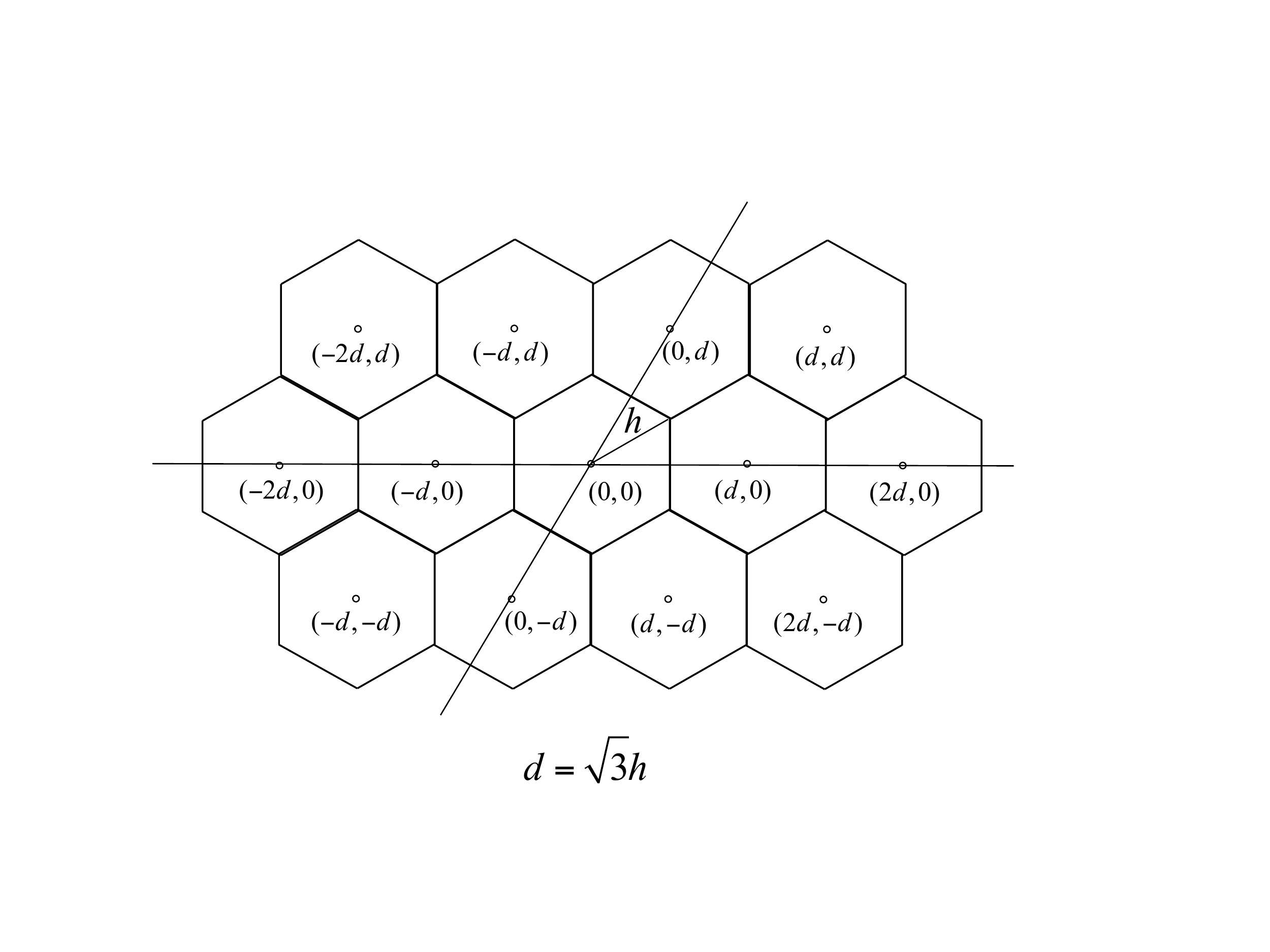}

Figure 2:  The hexagonal quantization of the transceiver coordinates
\label{fig2}
\end{center}
\end{figure}

A fundamental assumption of our model is:

\noindent
{\bf The Spatial Coordinate Quantization Assumption:}
There is a well defined mapping taking the GNSS location information into quantization cells.
Any user can determine at the local start time of a superframe the identity of the cell to which it belongs.

Although there is no upper bound on the number of users in the network, there is a bound on the number of users within
a hearing area.   We refer to this bound as the {\em maximum number of local users.}
In particular, we adopt:

\noindent
{\bf The Maximum Interferer Assumption:}
At any time and at any location there is no more than $M$ users in any area of size $\pi R^2$.
So the maximum number of interferers is $M-1$.

We are now ready to define the problem to be addressed in this paper.

\begin{definition}
Consider a multi-hop ad hoc network that satisfies all the assumptions stated in this section.
The network is said to provide block-free service with parameters $(M,L)$ 
if it employs transmission sequences with period $L$ such that
when the maximum number of local users everywhere is not more than $M$,
any user in the network can receive at least one contention-free transmission 
in each normal frame of a superframe from any user who is located within its hearing distance at the beginning of that superframe.   
{\rm (Note that by the Interference Assumption, the interference relations among users
are unchanged within a superframe, so users within hearing range will remain
so in the whole superframe.)}
\end{definition}

If a network can provide block-free service with parameters $(M,L)$ it can also guarantee that any user can transmit at least one contention-free packet in each of normal frame of a superframe to any user who is located within its hearing distance when that superframe starts.   

Such a definition of service is appropriate for applications in which total system throughput is not a primary concern, but instead
the focus is on how to guarantee all users can connect to their neighbors within time constraint.  That is, it is important to
ensure all user can transmit to and hear from its neighbors within a bounded time delay.  Such a guarantee cannot be provided under probabilistic based protocols, such as CSMA.

In subsequent sections we examine the question of how to design and allocate transmission sequences
under different synchronization assumptions.  One basic objective is to find sequences with the shortest sequence length.
We will only focus on solutions that do not require information on current transmission status, hence algorithms
that require users to monitor the channel before transmission are excluded.
Adding such a step could enhance performance, although at extra cost.

\section{RELEVANT CONCEPTS FROM PROTOCOL SEQUENCES}\setcounter{equation}{0}

We review in this section coding theory and protocol sequence results that would be needed in subsequent sections.  First of all, we specify the idea of a protocol sequence set.  

\begin{definition}
A protocol sequence set $\cal{P}$ consists of binary sequences having a common period $n$.
The weight of an element $X \in \cal{P}$ is the number of ``1"s in it and is denoted by $w(X)$.
\end{definition}

Define the rightward cyclic shift operation, $\cal R$, on a sequence $X$ by:
\begin{equation}
{\cal R}(X(1), X(2), \ldots, X(n)) = (X(n), X(1), \ldots, X(n-1)),
\end{equation}
where $X(i)$ denotes the $i$-th component of $X$.
By a slight abuse of notation, the same ${\cal R}$ represents the rightward cyclic shift operation on sequences
with different lengths.
The Hamming cross-correlation function of elements $X$ and $Y$ can be expressed as
\begin{equation}
H(X, Y)(\tau) = \sum_{i=0}^{n-1} X(i){\cal R}^{\tau}Y(i).
\end{equation}

A cyclically permutable code is defined to be a binary block code with length $n$ such that
all codewords have $n$ distinct cyclic shifts and all codewords are cyclically distinct.
For such codes, it is useful to extend the minimum Hamming distance to {\it cyclic minimum distance}, $d_c$,
which is the minimum Hamming distance between all cyclically shifted versions of any two codewords.
That is
\begin{equation}
d_c = \min_{X,Y} \min_{i} d(X, {\cal R}^i Y),
\end{equation}
where $d$ is the regular distance function between two codewords.

In \cite{Nguyen92} several methods for constructing cyclically permutable codes were proposed.
One of the approaches, which is based on Reed-Solomon codes, is relevant to our discussion.
Let $p$ be a prime number.  Consider a Reed-Solomon code, $C(n, p, k)$, which has codewords of length $n$ over GF($p$).
For $0 \leq i < p$ let $E_i(p)$ represent the sequence
\begin{equation}
E_i(p) = (\overbrace{\underbrace{0, \ldots, 0}_{i}, 1, 0, \ldots, 0)}^p.
\end{equation}

Denote the set $\{ E_1(p), \ldots, E_p(p) \}$ by ${\cal{E}}(p)$.
The $v_p$-representation is a mapping from GF($p$) to ${\cal{E}}(p)$ that maps $j$ to $E_j(p)$.
A codeword from $C(n, p, k)$ can then be viewed as a $p$-by-$n$ binary matrix.
Moreover, it is shown in \cite{Nguyen92} that this construction
defines a cyclically permutable code with codeword
length $pn$, constant weight $n$, codeword size $p^{(k-2)}$, and minimum distance $2(n-k+1)$,
for $3 \leq k < n \leq p$.

In the above result, the connection going from a $p$-by-$n$ matrix to a length $pn$ sequence is based on the Chinese Remainder Theorem (CRT) correspondence, (denoted by
$\gamma$,) which is a
mapping taking entries of a $cd$-dimensional vector to entries of a $c$ by $d$ matrix.
If $c$ and $d$ are relatively prime, the CRT implies that
such a correspondence is one-to-one and onto.   Moreover, it is proved
in \cite{Nguyen92} that a cyclic rightward shift of a $cd$-dimensional vector
is equivalent to a cyclic downward row shift  followed by a cyclic rightward
column shift of the corresponding matrix.  

The CRT correspondence has also been employed to construct several classes of protocol sequences, known as CRT sequences.
There are several variants of CRT sequences.
We recall the definition of a class of CRT sequences and their basic properties from \cite{SETA10} to facilitate subsequent discussions.

\begin{definition}
Let $p$ and $q \ge 2p-1$ be relatively prime positive integers.  A CRT sequence set, ${\cal{C}}(p,q)$, consists
of $p$ binary sequences with period $pq$ such that each element in it has a characteristic
set, (the set of positions in the sequence with value 1,) that can be summarized as:
\begin{equation}
{\cal{I}}_g = \left \{ l \in {\mathbb{Z}} : 0 \leq l < pq, \gamma(l) = (jg ~ {\rm mod} ~p, j), j=0, 1, \ldots, p-1 \right \},
\end{equation}
for $0 \leq g < p$.  The parameter $g$ is referred to as the sequence generator.
\end{definition}

Each of the $p$ distinct sequences in the set has a Hamming weight $p$.
Moreover, the Hamming cross-correlation between any shifted versions of any two distinct
sequences is either 0 or 1, which implies that the sequence set possesses the {\em User Irrepressible} (UI) property.
This terminology was initiated in \cite{Wong07, SWSC} and was given the following mathematical definition in \cite{CAC10}:

\begin{definition}
Consider a protocol sequence set with $k$ elements, each having a period $l$.
Each element is represented by a shifted version that is obtained by applying the operator $\cal R$ independently
for an arbitrary number of times.  Denote by $\bf M$ the $k \times l$ matrix obtained by
stacking these representations one above the other.  The protocol sequence set is User-Irrepressible (UI) if we can always
find a $k \times k$ submatrix of $\bf M$ which is a permutation matrix, regardless
of the number of shifts applied to produce the representations.
\end{definition}
The relevancy of UI to a slot-synchronized network can be apprehended by interpreting $k$ as the number of users and the protocol sequences
as transmission sequences.  The UI condition is equivalent to the requirement that no matter what the relative shifts are, each users are guaranteed to have one conflict-free transmission in each period of $l$ slots if propagation delays are ignored.
Moreover, we use the phrase, a conflict-free ``1",
to refer to a sequence entry which corresponds to a column with only one non-zero entry
in $\bf M$.

The CRT sequence sets are among the shortest known UI sequence sets.
In \cite{SETA10} there is a table summarizing the shortest UI sequences sets for small user number.
In it, many of the known cases are CRT type sequences. 

We conclude this section with  a slightly modified version of a CRT sequence set for our subsequent discussion.

\begin{definition}
Let $p$ and $q \ge 2p-1$ be relatively prime positive integers.
A ${\cal{C}}_0  (p,q)$ sequence set consists
of $p$ binary sequences with period $pq$ of the following types:

1. For $g$ in $\{0\} \cup \{2, \ldots, p \}$, there are $p-1$ CRT sequences, $S_g$, with characteristic set
${\cal{I}}_g$.

2. A single sequence, $S_*$, with characteristic set:
\begin{equation}
{\cal{I}}_* = \left \{ l \in {\mathbb{Z}} : 0 \leq l < pq, \gamma(l) = (j,0), j=0, 1, \ldots, p-1 \right \},
\end{equation}

\end{definition}

It follows from results in \cite{SETA10} that the following holds.

\begin{lemma}
If $q \geq 2p-1$, the Hamming cross-correlation function of any two distinct sequences in ${\cal{C}}_0  (p,q)$ is upper bounded by {\rm 1}.
\end{lemma} 

\section{THE BASICS OF GNSS-BASED CODE ALLOCATION}\setcounter{equation}{0}
In this section we describe how transmission sequences are distributed to the users,
assuming appropriate sequences have already been identified.
In traditional code-based scheduling models,
transmission sequences are pre-assigned to users.
For applications such as VANETs, since all cars in a country may be potential users,
the dimension of the required code-space could be extreme large, even though
it is unlikely that any two users would interfere with each other.
In fact, in this paper we assume the number of users to be theoretically unbounded.  However, the number of
users within a geographical region is bounded and likely to be small.   So, we propose an alternative
sequence distribution scheme that is based on the instantaneous geographical location of the users.
In this section, we describe in detail how users can acquire transmission sequences automatically based on their geographical
locations.

Recall that under our stated assumptions, all users acquire a unique cell identity number at the
start time of their local superframes.  The transmission sequence will be based on this cell identity.
To avoid duplications in the sequence assignment, it is necessary that no two users will acquire the same cell identity at the
same superframe.  (The perceived start times of the superframes may differ as explained in Section 2.)

If all users are perfectly clock synchronized, this requirement translates to a condition that no two users can occupy
the same quantization cell at the same time.  Since the quantization cells are small, this is a mild condition to impose.
For the general case where users are not perfectly synchronized, 
let $\tau_i (k)$ represent the start time of superframe $k$ for user $i$ and let $(x_i {(k)}, y_i {(k)})$ represent the quantization
cell the user is located at $\tau_i (k)$. 
To maintain unique code sequence assignment, we require the follow condition holds:

\noindent
{\bf The Fermion Condition:}
For any superframe $k$, if
$(x_i {(k)}, y_i {(k)}) = (x_j {(k)}, y_j {(k)})$, then $i=j$.

If $\tau_i (k) = \tau_j (k)$, the Fermion Condition just requires the users cannot occupy the same quantization cell at the same time.
Since $\tau_i (k)$ and $\tau_j (k)$ are usually different,
this condition basically requires that the users need to be further separated than the synchronized cases
so that the maximum distance traveled during the time gap cannot affect the uniqueness of the
quantization.  For example, consider a case where $h=0.5$ m,
$\Delta \tau=100$ ms, and $v=60$ km/hr, then as long as the users are not closer than $4.34$ m
at any time, the Fermion Condition will hold.  For vehicular ad hoc network applications this is not a severe restriction.

The transmission sequence allocation is defined by a mapping, $\alpha$, from
$\cal{C}$, the set of cell identities, to $\cal{L}$, the set of defined transmission sequences.
It is assumed that this mapping is downloaded to all users at the time when they join the network
and can be updated from time to time, but not dynamically.
Since $\cal{C}$ is theoretically an unbounded set and $\cal{L}$ is finite, it
is necessary to reuse some of the sequences in $\cal{L}$.   As in the solution to frequency channel allocation, we address this issue by defining
a basic tiling pattern, in other words a cell cluster, and impose the condition that $\alpha$ maps cells in the cluster to distinct sequences.
The tiling pattern is then repeated to cover the whole area and thereby extending $\alpha$ to all of $\cal{C}$.
It follows that $\alpha$ can be computed easily with finite amount of data storage.

Under the Interference Assumption, users can hear each other only if their distance separation is less than $R$
at the beginning of a superframe.  However, due to the hidden node problem, a receiver may hear interfering signals from multiple transmitters separated by a distance less than $2R$.

As a result, it is permissible only for users separated by a distance greater than or equal to $2R$
to share the same transmission sequence and that users separated by a distance less than $2R$ should not use the same
transmission sequence.   Since the quantization cells are small compared to $R$, there is little error in assuming
the users are always located at the cell centers.  This assumption enables us to simplify the allocation constraint as:

\noindent
{\bf The Sequence Allocation Constraint:}
{\em Cells separated by a distance less than $2R$ cannot share the same transmission sequence.}

In our allocation algorithm design we will focus on finding solutions that satisfy this constraint.
Fortunately, this problem has already been solved in the context of frequency division multiple access.
Tilings pattern can be determined to satisfy this constraint according to the algorithm in \cite{Mac}.
For completeness, we will briefly summarize the procedure here.  Let $G$ be the smallest integer
that satisfies the following condition:
\begin{equation}
b_1^2+b_1b_2+b_2^2 =  G \geq \left( \frac{2R}{d} \right)^2,
\label{eqn4.2}
\end{equation}
for some non-negative integers, $b_1$ and $b_2$.

If a cell centered at $(m_1, m_2)d$ is assigned a sequence $S_1$, we assign the same sequence to all cells centered at:
\begin{equation}
(m_1+ib_1, m_2+jb_2)d.
\end{equation}
Here, $i$ and $j$ are arbitrary integers.  If there is another cell centered at a distance less than $2R$ from $(m_1, m_2)d$,
a different sequence other than $S_1$, say $S_2$, should be assigned.   Then, $S_2$ can be assigned to a lattice of cells
in a way similarly to $S_1$.   This allocation process is repeated until all cells are assigned a sequence.

It is proven in \cite{Mac} that the minimal number of sequences required to fill the whole space
so that no cells with distance $2R$ are assigned the same sequence is equal to $G$.  For example,
consider a case where the quantization radius is $h=1~m$ and $R=0.5~km$, then $G=333,333$.   This is not a
small number but it can be a couple of orders less than the total number of users in a system.   For example, the number of
registered vehicles in a country could be in the order of several tens of millions.

In subsequent sections we will consider how to construct enough sequences so that all
users can be guaranteed the user irrepressible property.  We also want to achieve this with short frame period.
While there are different methods for constructing UI sequences,
a common theme is that the mapping taking a
cell in a cluster to a sequence can be arbitrarily defined as long as it satisfies the one-to-one condition.
This approach greatly simplifies the transmission sequence design.

\section{PERFORMANCE OF GNSS-BASED SOLUTIONS}\setcounter{equation}{0}
Before discussing solutions based on protocol sequences, we present
several baseline models for comparison.  The first is TDMA scheduling. 

Assume that the system is perfectly synchronized, that is, $ \Delta = 0$.
For TDMA scheduling, we assign a slot to each cell in a cell cluster.  Thus, the minimum frame size is $G$, the value of which is defined in Equation (\ref{eqn4.2}).  Hence,
\[
L=G.
\]
When propagation delay and clock differences are taken into account, that is, when
$\Delta \ne 0$, there is no simple way to guarantee UI property
to all users except by appending to every time slot $\Delta$ silent slots in which no transmission is allowed.
These extra time slots allow propagation delay and clock difference effects be isolated from one transmission
to the next.  Hence, the minimum frame size is
\begin{equation}
L=(\Delta+1) G.
\end{equation}

To appreciate the physical significance of this construction, consider an example where
$h=1$ m, $R=0.5$ km, and $\Delta = 10$.  According to equation (\ref{eqn4.2})
this implies that the minimum period of a TDMA time frame
in which each user can be guaranteed to enjoy at least one data burst conflict-free
is 3,333,333.  In most applications, it is unlikely the number
of users within a radius of half a kilometer would exceed tens of thousands.  Thus, the TDMA scheme is extremely inefficient for this application.  Generally speaking, unless $M$ is in the order of $G$, TDMA is not an efficient choice.

If a sequence set with constant weight $w$ is used, in order to achieve the UI property it is clear that:
\begin{equation}
w \geq M,
\end{equation}
and
\begin{equation}
L \geq w \geq M.
\end{equation}
This bound is believed to be optimistic.
It is shown in \cite{SETA10} that for slot-synchronized cases, given any UI protocol sequence set that supports $M$ users,
the sequence length, $L(M)$, must satisfy
\begin{equation}
L(M) \geq \left \lceil \frac{8M^2}{9} \right \rceil.
\end{equation}
That is, the lower bound is quadratic in $M$.
Below we identify two sequence sets that satisfy the quadratic growth rate.

For the perfectly synchronized cases, there is a host of solutions in the literature under the heading of code-base scheduling,
Topology-Transparent Scheduling or Time Spread Multiple Access (TSMA), 
(see for example \cite{CF94}, \cite{JL98}, \cite{SCL03}, \cite{BB00}, \cite{RK2}, and \cite{RK3}.)
As explained previously, these approaches allocate the transmission sequences directly to the users rather than to the cells.
If $N$ denotes the number of users in the network, then it is shown in \cite{CF94} that one can construct a sequence set with period $L$
satisfying the order condition:
\begin{equation}
L=O \left( \frac{M^2ln^2N}{ln^2M} \right).
\label{eqn5.1}
\end{equation} 
In \cite{RK2}, a solution based on a $C(n, q, k)$ RS code has been proposed which has the same order property.

If we adopt these sequences to a GNSS-based system and assign them to cells in a cluster, the network can provide block-free service with appropriate parameters.  

\begin{proposition}
For a network under partial synchronization up to $\Delta$, a code-based scheduling
can provide block-free service with parameters $(M, L)$ where
\begin{equation}
L=O \left( (\Delta+1)  \frac{M^2ln^2G}{ln^2M} \right),
\end{equation} 
where $G$ is the number of cells in a cell cluster.
\label{prop1}
\end{proposition}
\noindent
{Proof}: Consider an arbitrary user, $u$, located at a quantization cell with center $O$ at time $S_k$.
Users who are located in cells with centers at a distance less than $R$ from $O$ at time $S_k$
are the only users who can be heard by $u$ during the superframe according to the Interference Assumption.
By the Maximum Interferer Assumption, the total number of such users including $u$ is at most $M$.
There exists a $C(n, p, k)$ RS code satisfying:
\begin{equation}
\begin{array}{c}
p^{k} \geq G,\\
n  \geq (k-1)(M-1)+1,\\
p  \geq  M.
\end{array}
\end{equation}
Moreover, the period of the corresponding transmission sequences is $np$.
By the construction of the cell cluster, all these $M$ users will be assigned distinct transmission sequences,
since the distance between any two of them is less than $2R$. 
If $\Delta = 0$, these conditions imply all users can transmit at least one conflict-free packet during a frame of $np$
slots.  This implies $u$ can receive at least one packet from each of the $M-1$ users during a normal frame.   
Following the same argument in \cite{CF94}, it is easy to show that
\begin{equation}
np=O \left(  \frac{M^2ln^2G}{ln^2M} \right).
\end{equation} 
For the partially synchronized case,  we can maintain the UI property if
we append every time slot in the RS code sequence by $\Delta$ silent slots in which no transmission is allowed.  So that
\begin{equation}
 L = (\Delta +1)np.
\end{equation}
$\hfill \blacksquare$

Even when $\Delta$ is of moderate value, the code-based scheduling approach could be impractical.  However, one can adopt
shorter solutions based on protocol sequences or cyclically permutable codes.
As mentioned in Section 3, there exists cyclic permutable codes with length $pn$, constant weight $n$, codeword size $p^{k-2}$,
and cyclic minimum distance $d_c \geq 2(n-k+1)$ for $3 \leq k < n \leq p$.
We can use this code to provide a solution to the asynchronous case by assigning distinct codewords to each cell
within a tiling pattern.   Hence,
\begin{equation}
\begin{array}{c}
p^{k-2} \geq G,\\
n  \geq (k-1)(M-1)+1,\\
p \geq  M.
\end{array}
\label{eqn5.2}
\end{equation}

\begin{proposition}
For the asynchronous case, it is possible to provide $(M, L)$ block-free service using cyclically permutable codes with
the period satisfying
\begin{equation}
L = O \left( \frac{M^2 \ln^2 G}{\ln^2 M} \right).
\end{equation}
\end{proposition}

\noindent
{Proof}:
The proof for the slot-synchronized case is similar to the proof of Proposition \ref{prop1}.
The cyclically permutable property implies that the UI property is maintained no matter what the relative shifts of
the sequences are.  For the asynchronous case, the transmission of a user may lie in between two consecutive
slots of the receiver.  However, as suggested in \cite{Massey82} if we append to every slot of the code sequence an extra silent slot in which
no transmission is allowed, the UI property will be preserved.   It follows that
\begin{equation}
L=2np.
\end{equation}
Using arguments similar to \cite{CF94}, the estimate for $L$ holds.
$\hfill \blacksquare$

\section{LOCATION-BASED PROTOCOL SEQUENCE DESIGN}\setcounter{equation}{0}

Comparing the GNSS-based allocation with the traditional code-based scheduling, an obvious
difference is that GNSS-based allocation does not set any {\em a priori} limit on the number of users in
the system.  Another important difference is that GNSS-based allocation offers the possibility for the allocation be dependent on geographic location.  For example, user
density clearly depends heavily on location; the number of users per unit area is higher in
downtown districts than in the countryside.  Allocation schemes that are designed based on
the maximum number of interferers seen in a downtown district may prove to be too pessimistic in the countryside. 

The protocol sequence sets we introduce so far are designed based on the assumption that the
maximum number of interferers, $M-1$, is constant in the whole geographic area. 
In this section, we consider the issue of designing protocol sequence sets that
allow variations on the maximum of number of interferers while maintaining the UI property.   Our proposed
solution is based on the idea of sharing a protocol sequence among multiple users.  
The sharing is enabled by means of {\em product sequences}.
These objects are distinct from product codes,  in spite of some superficial
similarities, (see \cite{Justesen} for a definition of product codes.)

Consider $\cal{P}$ and $\cal{Q}$, two protocol sequence sets with sequence length $p$ and $q$ respectively.
Let $X \in \cal{P}$ and $Y \in \cal{Q}$.
\begin{definition}
Suppose $p$ and $q$ are relatively prime,
the product sequence of $X$ and $Y$, denoted by $X \otimes Y$, is defined as follows.
Let ${\bf M}_{X, Y}$ be the $p$ by $q$ matrix with the $(i,j)$ entry defined by:
\begin{equation}
{\bf M}_{X,Y}(i,j) = X(i)Y(j).
\end{equation}
$X \otimes Y$ is the unique binary sequence that corresponds to ${\bf M}_{X,Y}$ through the CRT correspondence.
\end{definition}

It is clear that the length of $X \otimes Y$ is $pq$ and it has weight equal to $w(X)w(Y)$.

\begin{lemma}
The following properties hold for product sequences:
\begin{enumerate}
\item If $X \otimes Y = X' \otimes Y'$ is not the all zero sequence, then
\[
X=X', ~~~Y=Y'.
\]
\item ${\cal R}(X \otimes Y) = {\cal R}(X) \otimes {\cal R}(Y).$
\end{enumerate}
\end{lemma}

The proofs of these results are straightforward and are omitted.

Consider a cyclically permutable protocol sequence set, $\cal P$, such as the one constructed in Section 5.
Recall that we can construct it based on a $C(n, q, k)$ RS code so that the cyclic minimum distance satisfies
$d_c \geq 2(n-k+1)$.  For simplicity, we assume that $n$ and $q$ are primes.  Moreover,
\begin{equation}
\begin{array}{c}
n  \geq (k-1)(M-1)+1.
\label{eqn6.1}
\end{array}
\end{equation}

 It follows that any subset of $\cal P$ with $M$ sequences is UI.
We now show how to expand $\cal P$ to a larger protocol sequence
set, $\widetilde{\cal P}$, to accommodate more users while maintaining the UI property.
Let
\begin{equation}
L=nq
\end{equation}
represent the sequence length of elements in $\cal P$.
Let $p$ be a prime number such that $p \leq M$.
Let ${\cal Q} = \{S_1, \ldots, S_{p} \}$ be a subset of $\cal P$ and ${\cal Q}^C$ be its complement in $\cal P$.
 
The basic idea of the construction is that sequences in $\cal Q$
will be split to support additional users, while sequences in ${\cal Q}^C$ are
unchanged.  Regions using split codes can allow for more interfering users without
affecting nearby regions where the original codes from ${\cal Q}^C$ are used.

In mathematical notation, this can be formalized as follows.
Let $U$ represent the all ``1" sequence:
\begin{equation}
(\underbrace{1, 1, \ldots, 1}_{p(2p-1)}).
\end{equation}

For any sequence $X$ in ${\cal Q}^C$, the product sequence $U \otimes X$ repeats $p(2p-1)$ times the sequence $X$ and hence
its application essentially has the same effect as $X$ if the superframe is long enough.
Let
\begin{equation}
\widetilde{P}_1 = \{ U \otimes X: X \in {\cal Q}^C \}.
\end{equation}

Label elements in the CRT sequence set, ${\cal{C}}_0  (p,2p-1)$, by $\{ C_1, \ldots, C_p \}$.
Define the set $\widetilde{P}_2$ as follows:
\begin{equation}
\widetilde{P}_2 = \{ C_i \otimes S_i : 1 \leq i \leq p \}.
\end{equation}
Strictly speaking, the set $\widetilde{P}_2$ depends on how elements in the corresponding sets are labeled.   But for the results presented below, the ordering bears no important consequence and any ordering could be used.
The expanded set, $\widetilde{\cal P}$, is
defined to be the union of $\widetilde{P}_1$ and $\widetilde{P}_2$.  Note that sequences in $\widetilde{\cal P}$
has sequence length equal to $p(2p-1)nq$.   Those belonging to $\widetilde{P}_1$ have weight $p(2p-1)n$ and
those belonging to $\widetilde{P}_2$ have weight $pn$.

\begin{theorem}
Any subset of $\widetilde{P}$ that includes all elements of $\widetilde{P}_2$ and $M-1$ elements from $\widetilde{P}_1$
is UI.  Moreover, each sequence from $\widetilde{P}_1$ has at least
\begin{equation}
p(2p-1) - p(p-1)/2 = p(3p-1)/2
\end{equation}
conflict-free {\rm ``1"}s in each period.
\label{thm6.1}
\end{theorem}
\noindent
{Proof:}
Let $S=C_i \otimes S_i$ be a protocol sequence from $\widetilde{P}_2$.
Since $C_i$ is an element from the UI protocol sequence set, $C_0(p, 2p-1)$, it has at least one conflict-free ``1"
no matter how the sequences are shifted by $\cal R$.   This means in the matrix representation, $S$ has one row which does
not conflict with any other sequences from $\widetilde{P}_2$.   Since there are at most $M-1$ elements
from $\widetilde{P}_1$, there is at least one conflict-free ``1" entry in that row for $S$.

Now suppose the subset contains a sequence $S$ which belongs to $\widetilde{P}_1$.
Let ${\bf M}(S)$ be the matrix representation of $S$
and consider an arbitrary row, say row $r$ of ${\bf M}(S)$.   There are at most $M-2$ other elements
from $\widetilde{P}_1$ in the sequence subset.  Hence at most $(M-2)(k-1)$ of the ``1"s in row $r$ of $S$ conflict
with those coming from sequences that belong to $\widetilde{P}_1$.   
Therefore, the number of ``1" entries of $S$ in row $r$ that do not conflict with ``1"s from other sequences belonging to
$\widetilde{P}_1$ is at least:
\begin{equation}
n-(M-2)(k-1) \geq k.
\end{equation}
Since any two sequences in ${\cal Q}^C$ can have at most $k-1$ bits in conflict,
$S$ will have at least one conflict-free entry in row $r$ if there is no more than one sequence from $\widetilde{P}_2$ that has non-zero entries in row $r$ in its matrix representation. 
However, as $C_0(p, 2p-1)$ is a UI protocol sequence set, there cannot be more than $p(p-1)/2$ places where elements from $C_0(p, 2p-1)$ conflict with one another.
So $S$ has at least $p(2p-1) - p(p-1)/2$ conflict-free ``1" entries.
$\hfill \blacksquare$

Theorem \ref{thm6.1} implies that on the average the effect on the non-split sequences is not too severe if one sequence
is shared among up to $M$ users.  By further analysis, we can show that for the non-split sequences, the delay between
conflict-free ``1" entries is bounded by
\[
2pL.
\]
To prove this result, we need to establish the following:

\begin{lemma}
For any sequence $S$ in $C_0(p, 2p-1)$,  the minimal separation between two consecutive {\rm ``1"}s is
at least $p$.  That is, there are at least $p-1$ {\rm ``0"}s between any two {\rm ``1"}s.
\label{lemma6.1}
\end{lemma}
\noindent
{Proof:}
It is easy to check that the claim holds for $S_0$ and $S_*$.
Consider $S_g$ with $2 \leq g < p$.  Recall that the characteristic set of $S_g$ is
\begin{equation}
{\cal{I}}_g = \left \{ l \in {\mathbb{Z}} : 0 \leq l < p(2p-1), \gamma(l) = (jg ~ {\rm mod} ~p, j), j=0, 1, \ldots, p-1 \right \}.
\end{equation}

Let $l^*$ be the unique solution satisfying the following equations. (Recall that $p$ is a prime.)
\begin{equation}
\left\{
\begin{array}{ll}
l \equiv 1 & {\rm mod} ~2p-1, \\
l \equiv g & {\rm mod} ~p, \\
0 \leq l < pq. &
\end{array}
\right.
\label{eqn6.8}
\end{equation}
Then ${\cal{I}}_g$ can be rewritten as:
\begin{equation}
\{ 0, l^*, 2l^* { ~mod~ } p(2p-1), \ldots, (p-1)l^*  {~mod~ } p(2p-1) \}.
\end{equation}
We can relabel this set as $\{ 0, a_1, \ldots, a_{p-1} \}$ with the additional condition that
\begin{equation}
0 < a_1 < \cdots < a_{p-1}.
\end{equation}
One can show by induction that the distance between any two entries in the
characteristic set of $S_g$ is not less than the distance between $a_1$ and 0 and
$a_{p-1}$ and $p(2p-1)$.  That is, the minimal distance is equal to:
\begin{equation}
\min (a_1, p(2p-1)-a_{p-1}).
\end{equation}
For $g \ne 1$,
\begin{equation}
lg \ne l ~ {\rm mod} ~p.
\end{equation}
Hence,  for $j=0, 1, \ldots, p-1$, there exists no solution to the equation:
\begin{equation}
(l, l) = (jg ~ {\rm mod} ~p, j),
\end{equation}
for $l=1, \ldots, p-1$.  Moreover, for $l$ in this range, $\gamma(l)=(l,l)$, so
\begin{equation}
\gamma(l) \ne (jg ~ {\rm mod} ~p, j).
\end{equation}
In other words,  $l =1, \ldots, p-1$ cannot appear in ${\cal{I}}_g$.  Similarly, for
$l =p(2p-2)+1, \ldots, p(2p-1)-1$,
\begin{equation}
\gamma(l) = (l-p(2p-2), l-(p-1)(2p-1)).
\end{equation}
Since
\begin{equation}
l-(p-1)(2p-1) \geq p(2p-2)+1-(p-1)(2p-1) = p,
\end{equation}
$l =p(2p-2)+1, \ldots, p(2p-1)-1$ cannot appear in ${\cal{I}}_g$.
So
\begin{equation}
\min (a_1, p(2p-1)-a_{p-1}) > p-1.
\end{equation}
$\hfill \blacksquare$

\begin{lemma}
Consider the protocol sequence set $C_0(p, 2p-1)$.
Let $\bf M$ be a $p$ by $p(2p-1)$ binary matrix obtained
by listing the sequences in the set as rows.  Let $\bf F$ be derived from $\bf M$
by applying the operator $\cal R$ to each row independently
for an arbitrary number of times.  In $\bf F$, there is at least one zero column within
any $2p$ consecutive columns.
\label{lemma6.2}
\end{lemma}
\noindent
{\it Proof:}
Without lost of generality assume that the first $2p$ columns of the matrix contain no zero column.
Since the separation between the ``1" entries in each row is at least $p$, the ``1" entries in the first 
$p$ columns must come from different sequences.   Similarly, the same holds for the $p+1$-th to $2p$-th
columns.   Hence, for each sequence in $C_0(p, 2p-1)$, $S_i$, we can identify two numbers
$(a_i, b_i)$, which indicate the positions of two of its ``1" entries, such that
\begin{equation}
\begin{array}{l}
1 \leq a_i \leq p, \\
p+1 \leq b_i \leq 2p, \\
a_i \ne a_j,  {\rm ~if~} i \ne j, \; b_i \ne b_j, {\rm ~if~} i \ne j.
\end{array}
\end{equation}
Therefore,
\begin{equation}
\sum_{i=1}^p (b_i - a_i) = \sum_{i=1}^{2p} i - \sum_{i=1}^p i =  p^2.
\end{equation}
On the other hand, Lemma \ref{lemma6.1} implies that:
\begin{equation}
 b_i - a_i \geq p.
\end{equation}
As $C_0(p, 2p-1)$ is a UI set, and since the difference between consecutive ``1" entries
in $S_0$ is exactly $p$, $ b_i - a_i$ is larger than $p$ for the other sequences.  Hence,
\begin{equation}
\sum_{i=1}^p (b_i - a_i) >  p^2.
\end{equation}
A contradiction.
$\hfill \blacksquare$

\begin{theorem}
Consider a subset of $\widetilde{P}$ that includes all elements of $\widetilde{P}_2$ and $M-1$ elements from $\widetilde{P}_1$.
Let each element of the subset be represented by a shifted version that is obtained by applying the operator $\cal R$ independently
for an arbitrary number of times.  Form a matrix by
stacking these representations one above the other. 
For each row corresponding to a sequence in $\widetilde{P}_1$, the gap between consecutive conflict-free ``1"s
in the matrix is at most:
\begin{equation}
2pL.
\end{equation}
\label{theorem6.2}
\end{theorem}
\noindent
{\it Proof:}
Let $S=U \otimes X$ be a sequence from $\widetilde{P}_1$.  Without lost of generality
we can assume in the matrix representation of $S$ the entry at the first row and first column
corresponds to a conflict-free ``1", and that for all sequences from $\widetilde{P}_2$, the
first row contains only value 0.   Let $k$ be the nearest row in which all sequences from $\widetilde{P}_2$
contains only value 0.   For $S$, the next conflict-free ``1" will occur no later than $l$, where
$l$ satisfies:
\begin{equation}
\left\{
\begin{array}{ll}
l \equiv 0 & {\rm mod} ~L, \\
l \equiv k & {\rm mod} ~p(2p-1), \\
0 \leq l < p(2p-1)L. &
\end{array}
\right.
\end{equation}
That is $l=kL$.   By Lemma \ref{lemma6.2}, $k \leq 2p$.
$\hfill \blacksquare$

The estimate in Theorem \ref{theorem6.2} is a relatively crude.   It is believed that a smaller bound is possible,
maybe with some modification in the sequence design.   The CRT sequence approach requires working with
relatively prime parameters.  It is possible to relax this requirement by using Generalized Prime Sequences
as described in \cite{WuYi1}.

\section{CONCLUSION}\setcounter{equation}{0}
In this paper we propose a new approach to the MAC design that is based critically on
information obtained from a GNSS.   The approach merges idea from code-based
scheduling and protocol sequence.  It allows for differences in time synchronization
among different users and enables the possibility of traffic load fine-tuning based on geographic locations.  The concept of spatial reuse for transmission sequences is believe
to be novel and provides a new perspective on addressing MAC problem. 
This is potentially a fertile direction for further investigation.

\bibliographystyle{IEEEtran}

\bibliography{Mobile}

\end{document}